\documentclass[useAMS,usenatbib]{mn2e} 
\usepackage{aas_macros}
\usepackage{graphics}
\usepackage{epsfig}  
\usepackage{natbib} 
\usepackage{float}
\newcommand{\Tab}[1]{Table~\ref{#1}}
\newcommand{\Sec}[1]{Section~\ref{#1}}
\newcommand{\Eq}[1]{Eq.~(\ref{#1})}
\newcommand{\Fig}[1]{Figure~\ref{#1}}
\newcommand{\amiga}{\texttt{AMIGA}}
\newcommand{\ahf}{\texttt{AHF}}
\newcommand{\mlapm}{\texttt{MLAPM}}
\newcommand{\mhf}{\texttt{MHF}}

\def\hkpc{$h^{-1}{\ }{\rm kpc}$}
\def\hMpc{$h^{-1}{\ }{\rm Mpc}$}
\def\hMsun{$h^{-1}{\ }{\rm M_{\odot}}$}

\def\nbody{$N$-body}
\def\LCDM{$\Lambda$CDM}

\def\vecC{\mbox{\bf C}}
\def\vecx{\mbox{\bf x}}
\def\vecp{\mbox{\bf p}}
\def\vech{\mbox{\bf h}}
\def\vecr{\mbox{\bf r}}

\def\vecg{\mbox{\bf g}}

\def\vecnabla{{\nabla}}

\def\be{\begin{equation}}
\def\ee{\end{equation}}
\def\ba{\begin{eqnarray}}

\def\ea{\end{eqnarray}}

\title[MONDian Structure Formation]
{Cosmological Structure Formation under MOND: a new numerical solver for Poisson's equation}
\author[Llinares et al.]
{Claudio Llinares$^{1}$, Alexander Knebe$^{1}$, HongSheng Zhao$^{2}$\\
$^{1}$Astrophysical Institute Potsdam, An der Sternwarte 16, Germany\\
$^{2}$SUPA, School of Physics and Astronomy, University of St. Andrews, North Haugh, St. Andrews, Fife, KY16 9SS, UK}

\begin{document}

\date{Submitted Version ...}

\pagerange{\pageref{firstpage}--\pageref{lastpage}} \pubyear{2008}

\maketitle

\label{firstpage}

\begin{abstract}
  We present a novel solver for an analogue to Poisson's equation in
  the framework of modified Newtonian dynamics (MOND). This equation
  is highly non-linear and hence standard codes based upon tree
  structures and/or FFT's in general are not applicable; one needs to
  defer to multi-grid relaxation techniques. After a detailed
  description of the necessary modifications to the cosmological
  \nbody\ code \amiga\ (formerly known as \mlapm) we utilize the new
  code to revisit the issue of cosmic structure formation under
  MOND. We find that the proper (numerical) integration of a MONDian
  Poisson's equation has some noticable effects on the final results
  when compared against simulations of the same kind
  but based upon rather ad-hoc assumptions about the properties
  of the MONDian force field. Namely, we find that the large-scale
  structure evolution is faster in our revised MOND model leading to
  an even stronger clustering of galaxies, especially when compared to
  the standard \LCDM\ paradigm.
\end{abstract}


\section{Introduction} \label{sec:introduction}

Modified Newtonian dynamics (MOND) was proposed by \citet{Milgrom83}
as an alternative to Newtonian gravity to explain galactic dynamics
without the need for dark matter. Although current cosmological
observations point to the existence of vast amounts of non-baryonic
dark matter in the Universe \citep[e.g.][]{Komatsu08}, it remains
interesting to explore other alternatives, especially as not all of
the features of CDM models appear to match observational data (e.g.,
the ``missing satellite problem'' \citep{Klypin99, Moore99} and the
so-called ``cusp-core crisis'' \citep[e.g.][]{deBlok03,Swaters03}). In
that regards it appears important to look for tests able to
discriminate between MOND and Newtonian gravity, especially in the
context of cosmology now that there exist various relativistic
formulation of the MOND theory
\citep{Bekenstein04,Zhao07,Zhao08}. However,
progress in that field has been hampered by the fact that MOND is a
non-linear theory making any analytical predictions as well as
numerical simulations a tedious task. To date, only a few codes exist
that actually solve the MONDian analogue to Poisson equation in a
non-cosmological context \citep{Brada99, Nipoti07, Tiret07}, none of
which is publically available. We are augmenting this list by making
available a new solver for the MONDian Poisson's equation primarily
designed to work in a cosmological context but readily adjusted to
allow for simulations of isolated galaxies.

Until recently MOND was merely a heuristic theory tailored to fit
rotation curves with little (if any) predictive power for cosmological
structure formation. One of the most severe problems for the general
appreciation and acknowledgment of MOND as a "real" theory (and a
conceivable replacement for dark matter) was the lack of success to
formulate the theory in a general relativistic manner.  This situation
though changed during the last couple of years and at present there
are a number of covariant theories (e.g. \citep{Bekenstein04,
Zhao07, Zhao08}) whose non-relativistic weak
acceleration limit accords with MOND while its non-relativistic strong
acceleration regime is Newtonian. However, there remains a lot to be
done with regards to the relativistic formulations of MOND in order to
a get completly acceptable theory. But nevertheless has MOND reached a
stage of development in wich we can think in doing cosmology with
reducing the need of unjustifiable assumptions.  Furthermore, the
study of cosmological structure formation in the non-linear regime
will provide new constrains on such generalizations of MOND.

The first valiant attempts at simulating cosmic structure formation
under the influence of MOND were done by \citet{Nusser02} and
\citet[][KG04 from now on]{Knebe04}. While their studies provided
great insights into the non-linear clustering behaviour of MONDian
objects (we dare to call them galaxies as none of these simulations
included the physics of baryons) they were still based upon some
rather unjustifyable assumptions. The main objective of this paper now
is to refine the implementation of MOND in the \nbody\ code \amiga\
(successor of \texttt{MLAPM}, \citet{Knebe01}), i.e. we modified the
code to numerically integrate a MONDian analogue to Poisson's
equation. This equation is a highly non-linear partial differential
equation whose solution is non-trivial to obtain. Only sophisticated
multi-grid relaxation techniques (on adaptive meshes of arbitrary
geometry) are capable of tackling this task. We argue that only when
MOND has been thoroughly and "properly" studied and tested against
\LCDM\ can we safely either rule it out for once and always or confirm
this rather venturesome theory. The theory has become a valid
competitor to dark matter and it therefore only appears natural -- if
not mandatory -- to (re-)consider its implications. 
We developed a
tool and the subsequently necessary analysis apparatus allowing to
test and discriminate cosmological structure formation in a MONDian
Universe from the standard dark matter paradigm and used it 
to study a particular model of MOND theory.

\section{The MONDian Equations} \label{sec:equations} Despite the
recent progress made in obtaining, refining and studying covariant
formulations of the MOND theory \citep[cf. ][]{Bekenstein04, Zhao07,
  Zhao08}, there are still certain ambiguities in
contriving a closed a set of equations-of-motion suitable for an
\nbody\ code. We will work out such a set in this Section.

\paragraph*{Non-MONDian Cosmology} 
We like to remind the reader that in a cosmological \nbody\ code one
integrates the \textit{comoving} equations of motion

\begin{equation}\label{eq:eom}
 \begin{array}{rcl}
  \displaystyle \dot{\vecx}          
   & = &\displaystyle  {\vecp \over m a^2} \ , \\ 
\\
  \displaystyle \dot{\vecp}          
   & = &-\displaystyle  m \nabla \Phi_N \\
 \end{array}
\end{equation}

\noindent
which are completed by Poisson's equation for the Newtonian potential
$\Phi_N$ responsible for peculiar accelerations

\begin{equation}\label{eq:PoissonNewton}
  \displaystyle \vecnabla \cdot \left[ \nabla\Phi_N(\vecx) \right]
  = \frac{4 \pi G}{a} \left(\rho(\vecx) - \overline{\rho} \right) \ .
\end{equation}

\noindent
In these equations $a$ is the cosmic expansion factor, $\vecx =
\vecr/a$ the comoving coordinate, $\vecp$ the canonical momentum,
$\vecnabla \cdot$ the divergence operator with respect to $\vecx$,
$\rho$ ($\bar \rho$) the comoving (background) density, and

\be \label{eq:ForceNewton}
\vecg_N(\vecx) = -\nabla \Phi_N(\vecx)
\ee

\noindent
the Newtonian \textit{peculiar acceleration field in comoving coordinates}.

The next step now is to define an analogon to \Eq{eq:PoissonNewton} that
takes into account the effects of MOND in cosmology.

\paragraph*{MONDian Cosmology} 
In order to get an equation for the comoving peculiar MONDian
potential $\Phi_M$ we have to make a decision about which covariant
generalization of MOND we want to use. Assuming the $\mu$ function to
be constant in space, a non-relativistic limit of the covariant theory
by \citet{Zhao08} can be written as follows

\be \label{eq:PoissonMOND} 
\nabla\cdot\left[
  \mu\left(\frac{|\nabla\Phi_{M}|}{a \gamma(a)}\right)\nabla\Phi_{M}\right]=\frac{4\pi G}{a}
\left(\rho-\bar\rho\right) \ . 
\ee

\noindent
where $\mu(x)$ is $\mu\rightarrow1$ for $x\gg 1$ (Newtonian limit) and
$\mu\rightarrow x$ for $x\ll 1$ (MONDian limit)
\citep[cf.][]{Milgrom83}. We further took the liberty to encode the
MONDian acceleration scale $\gamma(a)$ as a (possible) function of the
cosmic expansion factor $a$ for reasons that will become clear later
on (cf. \Sec{sec:cosmology}). The most naive choice would be
$\gamma(a) = g_0 = 1.2\times 10^{-8}{\rm cm}/{\rm sec}^2$ whereas
other theories may lead to different dependencies; for instance, in
\citet{Zhao08} $\gamma(a)$ is given as $\gamma(a)=a^{1/2}g_0$.

In analogy to \Eq{eq:ForceNewton} we define

\be \label{eq:ForceMOND}
\vecg_M(\vecx) = -\nabla \Phi_M(\vecx)
\ee

\noindent
as the MONDian \textit{peculiar acceleration field in comoving coordinates}.

\paragraph*{The Curl-Field} 
The relation between the MONDian and the Newtonian force,
i.e. \Eq{eq:ForceNewton} and \Eq{eq:ForceMOND}, is given by
\citep{Bekenstein84}

\be \label{eq:curl}
\mu\left(\frac{|\textbf{g}_M|}{g_0}\right)\textbf{g}_M = \textbf{g}_N
+  \mathbf{C} \ ,
\ee

\noindent
where an otherwise unspecified curl-field 

\be
\mathbf{C} = \nabla \times \mathbf{h} \ ,
\ee

\noindent
appears that has been shown to vanish for any kind of symmetry
\citep{Bekenstein84}. Previous studies on cosmological structure
formation under MOND neglected the curl-field $\mathbf C$ as this
allows to use a standard solver for the Newtonian Poisson's
equation~(\ref{eq:PoissonNewton}) and then an inversion of
\Eq{eq:curl} to obtain the MONDian forces \citep[][KG04]{Nusser02}.
However, this leads to parallel Newtonian and MONDian forces which is
not necessarily the case! The more correct approach is to directly
solve \Eq{eq:PoissonMOND} and in the following Section we present a
novel solver that numerically solves for the MONDian potential
$\Phi_M$ to be used with \Eq{eq:ForceMOND} in order to integrate the
equations-of-motion~(\ref{eq:eom}) for $N$ particles. The solution to
\Eq{eq:PoissonMOND} can in turn be used together with \Eq{eq:curl} to
actually study the curl-field $\mathbf C$ and its effects.

\section{The code} \label{sec:code}

Conventional Poisson solvers used throughout (computational) cosmology
are no longer applicable to \Eq{eq:PoissonMOND}. All of the standard
methods such as Fast-Fourier-Transform (FFT) based Particle Mesh (PM)
codes, tree codes, expansion codes, and their variants rely on the
linearity of Poisson's equation. The corresponding MONDian Poisson
equation~(\ref{eq:PoissonMOND}) though is a non-linear partial
differential equation which requires more subtle and refined
techniques to be solved numerically. We are now going to elaborate
upon the steps required to adjust a given relaxation solver to account
for the complexity of the MONDian Poisson's equation and any arbitray
MONDian interpolation function $\mu(x)$.

\subsection{Solving the MONDian Poisson's equation via multi-grid relaxation}
The code used is a modification of the open source cosmological code
\amiga\ that is the sucessor of \mlapm\ \citep{Knebe01}. We conserve
the equations of motion of the original code (i.e. \Eq{eq:eom}) but
we now numerically integrate the MONDian Poisson's
equation~(\ref{eq:PoissonMOND}). The fact that the new equation for the
potential is non-linear prevents us from using the standard methods
(cf. above); we rather need to adhere to a multi-grid relaxation
technique \citep[e.g.,][]{Brandt77, Press92, Knebe01}. The method
consists of discretizing the equation on a grid and solving the non
linear system of algebraic equations by relaxing a trial potential
$\Phi$ until convergence. We adhere to the discretisation proposed by
\citet{Brada99} and already used by \citet{Tiret07}

\ba \label{eq:discretiz}
\nonumber & &(\mu_{i+1/2}(\phi_{i+1} - \phi) +
\mu_{i-1/2}(\phi - \phi_{i-1}) +\\
\nonumber & & \mu_{j+1/2}(\phi_{j+1} - \phi) +
\mu_{j-1/2}(\phi - \phi_{j-1}) + \\
\nonumber & & \mu_{k+1/2}(\phi_{k+1} - \phi) +
\mu_{k-1/2}(\phi - \phi_{k-1}) ) / {h^2} = \\
& & \frac{4\pi G}{a}\rho_{i,j,k} ,
\ea

\noindent
where $\rho_{i,j,k}$ is the density contrast in cell $(i,j,k)$ on a
grid with spacing $h$, $\phi$ is the potential in cell $(i,j,k)$ with
$\phi_{i\pm 1}$ the potential in the neighbouring cells $(i\pm
1,j,k)$. The same terminology applies to the other two spatial
dimensions.  The coefficient $\mu_{i+1/2}$ is the evaluation of the
$\mu(x)$ function at the point $(i+1/2, j, k)$, etc.

\noindent
Reordering \Eq{eq:discretiz} we get

\ba \label{eq:disc_poi}
\nonumber & &\mu_{i+1/2}\phi_{i+1} +
\mu_{i-1/2}\phi_{i-1} +\\
\nonumber & &\mu_{j+1/2}\phi_{j+1} +
\mu_{j-1/2}\phi_{j-1} +\\
\nonumber & &\mu_{k+1/2}\phi_{k+1} +
\mu_{k-1/2}\phi_{k-1} -\\
\nonumber & &
(\mu_{i+1/2}+\mu_{i-1/2}+\mu_{j+1/2}+\\
\nonumber & & \mu_{j-1/2}+\mu_{k+1/2}+\mu_{k-1/2})\phi =\\
& & \frac{4\pi G}{a}\rho_{i,j,k}
\ea

\noindent
where we can see that this way of discretising the equation is similar to the
standard discretization of the Newtonian Poisson's equation, but the
potential is ``weighted'' by the $\mu$ function. Given the nature of the $\mu$
function under MOND it can be set to unity for accelerations larger than $g_0$
and we then recover the standard discretisation of the Newtonian Poisson's
equation (cf. equation~(11) in \citet{Knebe01}).

The gradient of the potential, i.e. $\nabla \phi$, in the argument of
$\mu(x)$ needs to be disctretized, too. We choose the following form

\ba \label{eq:gradient}
\left(          (\nabla\phi)_x\right)_{i+1/2} & = & \frac{\phi_{i+1} - \phi}{h} \\
\left(\nonumber (\nabla\phi)_y\right)_{i+1/2} & = & \frac{(\phi_{i+1,j+1} + \phi_{i+1})-(\phi_{i+1,j-1} + \phi_{j-1})}{4h},\\
\left(\nonumber  (\nabla\phi)_z\right)_{i+1/2} & = & \frac{(\phi_{i+1,k+1} + \phi_{k+1})-(\phi_{i+1,k-1} +
  \phi_{k-1})}{4h} 
\ea

\begin{figure}
  \begin{center}
    \includegraphics[width=0.45\textwidth]{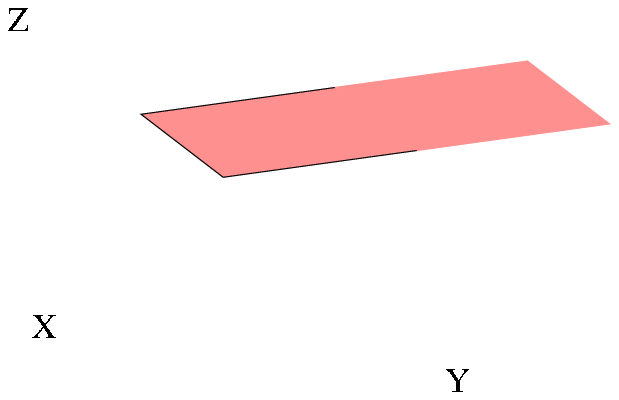}
    \caption{Stencil (27 points) used in the discretization of the
      MONDian Poisson's equation (cf. \Eq{eq:disc_poi}).  The argument
      of the $\mu(x)$ function is approximated in points indicated by
      the open circle, i.e. in-between the standard nodes.  The
      triangles show the cells used to calculate the gradient in that
      point (cf. \Eq{eq:gradient}).}
    \label{fig:discretiz}
  \end{center}
\end{figure}

\noindent
The nodes used in this discretisation are highlihted in
\Fig{fig:discretiz}.

There are two possible choices for the relaxation procedure. We can
either re-order terms in \Eq{eq:disc_poi} again (while freezing the
coefficients $\mu_{i\pm 1/2}, \mu_{j\pm 1/2}, \mu_{k\pm 1/2}$) in a
way that we get

\be \label{eq:iteration}
\phi = f(\phi_{i\pm 1}, \phi_{j\pm 1}, \phi_{k\pm 1}, \mu_{i\pm  1/2}, \mu_{j\pm 1/2}, \mu_{k\pm 1/2}, \rho_{i,j,k}) ,
\ee

\noindent
or we can apply one step of a Newton-Raphson root-finding algorithm to
$0 = \phi-f(\phi_{i\pm 1}, \phi_{j\pm 1}, \phi_{k\pm 1}, \mu_{i\pm
  1/2}, \mu_{j\pm 1/2}, \mu_{k\pm 1/2}, \rho_{i,j,k})$
\citep{Press92}. These two alternatives show similar rate of
convergence that goes in favour of one or the other according to the
density.  Our method of choice is the iterative procedure as it
involves fewer calculations.

\begin{figure}
  \begin{center}
    \includegraphics[width=0.5\textwidth]{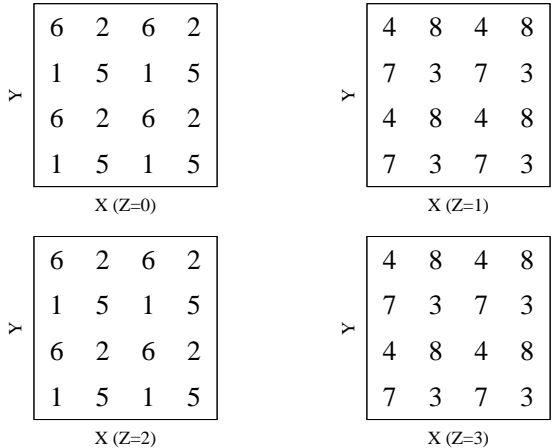}
    \caption{Eight-colour scheme used for obtaining an iterative
      solution of \Eq{eq:PoissonMOND} (cf. also \Eq{eq:iteration}).
      The numbers indicate the ordering by which the nodes that are
      updated.}
    \label{fig:colors}
  \end{center}
\end{figure}

We still need to define a ``colouring''-scheme, i.e. a way of how to
sweep through the whole grid and (iteratively) update the potential
$\phi$ in a given cell $(i,j,k)$.  The ordering of the iterations is
given by a generalization of the standard two-colour/red-black method
and sketched in \Fig{fig:colors}.  One iteration step is complete
after sweeping eight times trought the grid updating those nodes per
sweep that have identical numbers as indicated in \Fig{fig:colors}.

We further coded additional iteration schemes (e.g., lexicographic and
zebra in three directions) and the code switches automatically between
them in extreme cases for which the convergence with the 8 colours is
slow. For an elaborate discussion of multi-grid relaxation techniques
and colouring-schemes in particular we refer the interested reader to
\citet{Wesseling92}.

A numerical solution $\Phi^k_n$ has converged after $n$ iterations on a grid $k$ if
the norm $||\cdot||$ (mean or maximun value in the grid) of the
residual
\be \label{eq:residual}
{\it e}^k = L^k(\Phi^k_n)-\frac{4\pi G}{a} \left(\rho-\bar\rho\right)
\ee


\noindent
is small compared against the norm of the truncation error

\be \label{eq:truncation_error}
\tau^k = L^{k-1}(R\Phi^k_n) - R L^k(\Phi^k_n)
\ee

\noindent
i.e. $||{\it e}^k|| < 0.1 ||\tau^k||$. The value $0.1$ has been
determined heuristically.  Here, $L^k$ denotes the discretization of
the left hand side of \Eq{eq:PoissonMOND} on a grid $k$, $L^{k-1}$ the
same discretization in the next coarser grid and $R$ is the
restriction operator used to interpolate values from the grid $k$ to
the grid $k-1$.

\subsection{Tests} \label{sec:tests}
In order to verify and gauge the credibility of our numerical solver
we performed two complementary tests. The first simply assesses the
recovery of the potential for a given solution of the MONDian
Poisson's equation. The second test checks whether or not we recover
the same temporal evolution of a (cosmological) simulation when
setting the MONDian acceleration $g_0$ to such a low value that it
will not affect structure formation (Newtonian limit).

\subsubsection{Static test}
\begin{figure}
  \begin{center}
    \includegraphics[width=0.45\textwidth]{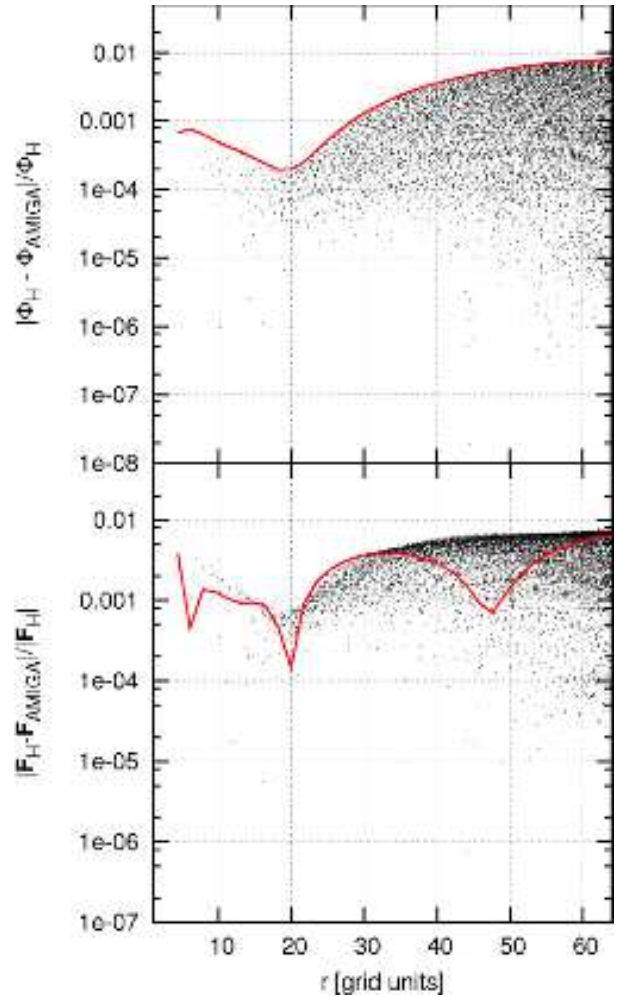}
    \caption{Relative error in the potential (upper panel) and the
      forces (lower panel) recovered with the solver on a $64^3$ grid
      for a Hernquist profile as a function of radius. The solid (red)
      line is the diagonal across the box while the dots represent a
      random sample of all grid points.}
    \label{fig:hernquist_errors}
  \end{center}
\end{figure}

As a first (static) test for the numerical potential solver, we
present the recovery of the analytical potential for a known
analytical solution of the MONDian Poisson's
equation~(\ref{eq:PoissonMOND}). To this extent we actually start with
an analytical description for a spherically symmetric potential akin
to the Hernquist profile \citep{Hernquist90}

\be \label{eq:HQSTpot}
\Phi_H = -\frac{GM}{r+r_c} + \sqrt{G M
  g_0}\ln\left(\frac{r+r_c}{r_0}\right) .
\ee 

\noindent
We added a logarithm to the potential in order to mach the properties
of the solution of the MOND equation (far from any source, the MONDian
forces are proportional to $\ln{(r)}$). We then derive the density by
substituting \Eq{eq:HQSTpot} into \Eq{eq:PoissonMOND}

\be
\begin{array}{rcl}
\rho & = & \displaystyle  \frac{1}{r^2}\left[\frac{2r}{g_0}\frac{A^2}{C} -
  \frac{2r^2}{g_0}\frac{A B}{C} +
  \frac{r^2}{g_0^2}\frac{A^2 B}{C^2}\right] \\
\\
A & = & \displaystyle \frac{G M}{(r+r_c)^2} + \frac{\sqrt{G M g_0}}{(r+r_c)} \\
\\
B & = & \displaystyle \frac{2 G M}{(r+r_c)^3} + \frac{\sqrt{G M g_0}}{(r+r_c)^2} \\
\\
C & = & \displaystyle 1+\frac{A}{g_0}
\end{array}
\ee

\noindent
The test is performed with non-periodic boundary conditions on a regular
64$^3$ grid by fixing the solution on the border to the analytical values.
\Fig{fig:hernquist_errors} now shows the fractional error in the potential
(upper panel) and force (lower panel).  The constants in the potential for
this particular test are: $M=10^{10}M_\odot$, $r_c=2.5\times 10^{-3}$ Mpc and
$r_0=0.01$ Mpc.  The box used was $0.01$ Mpc.  With this choice for the
parameters, the force gradually changes from a purely Newtonian regime 
in the
central parts to deep MONDian in the outer regions. The fractional errors
are calculated as the difference between the analytical potential and forces
on the grid and the numerical solution obtained by our solver.  The solid red
line shows the error as a function of radius along the diagonal in the box
while the dots represent a random sample of all grid points.\footnote{Note
  that running along the 3-dimensional diagonal will coincide with the largest
  possible error and hence the solid line in the upper panel of
  \Fig{fig:hernquist_errors} marks the upper boundary of the error in the
  potential.} We can clearly see that the error is never larger than 1~per
cent indicating the excellent quality of our numerical integration scheme.

\subsubsection{Dynamic test}
\begin{figure}
  \begin{center}
    \includegraphics[width=0.45\textwidth]{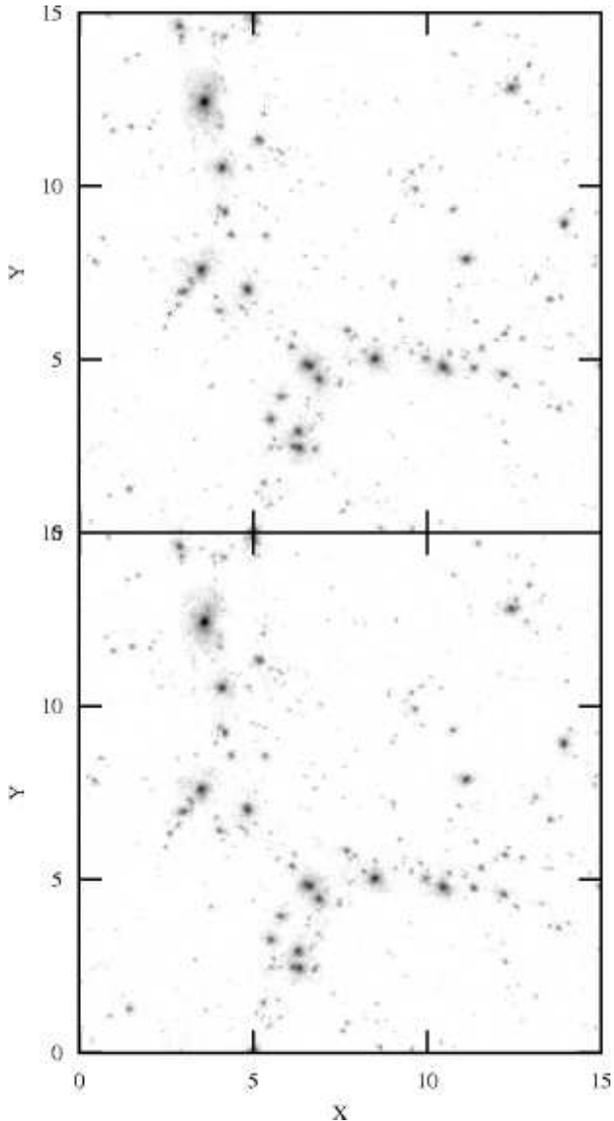}
    \caption{Density distribution at redshift $z=0$ for the $64^3$
      particle test simulation in the Newtonian limit ($g_0=10^{-12}$
      cm/s$^2$, upper panel) and the corresponding Newtonian run
      (lower panel).}
    \label{fig:density_064}
  \end{center}
\end{figure}

\begin{figure}
  \begin{center}
    \includegraphics[width=0.45\textwidth]{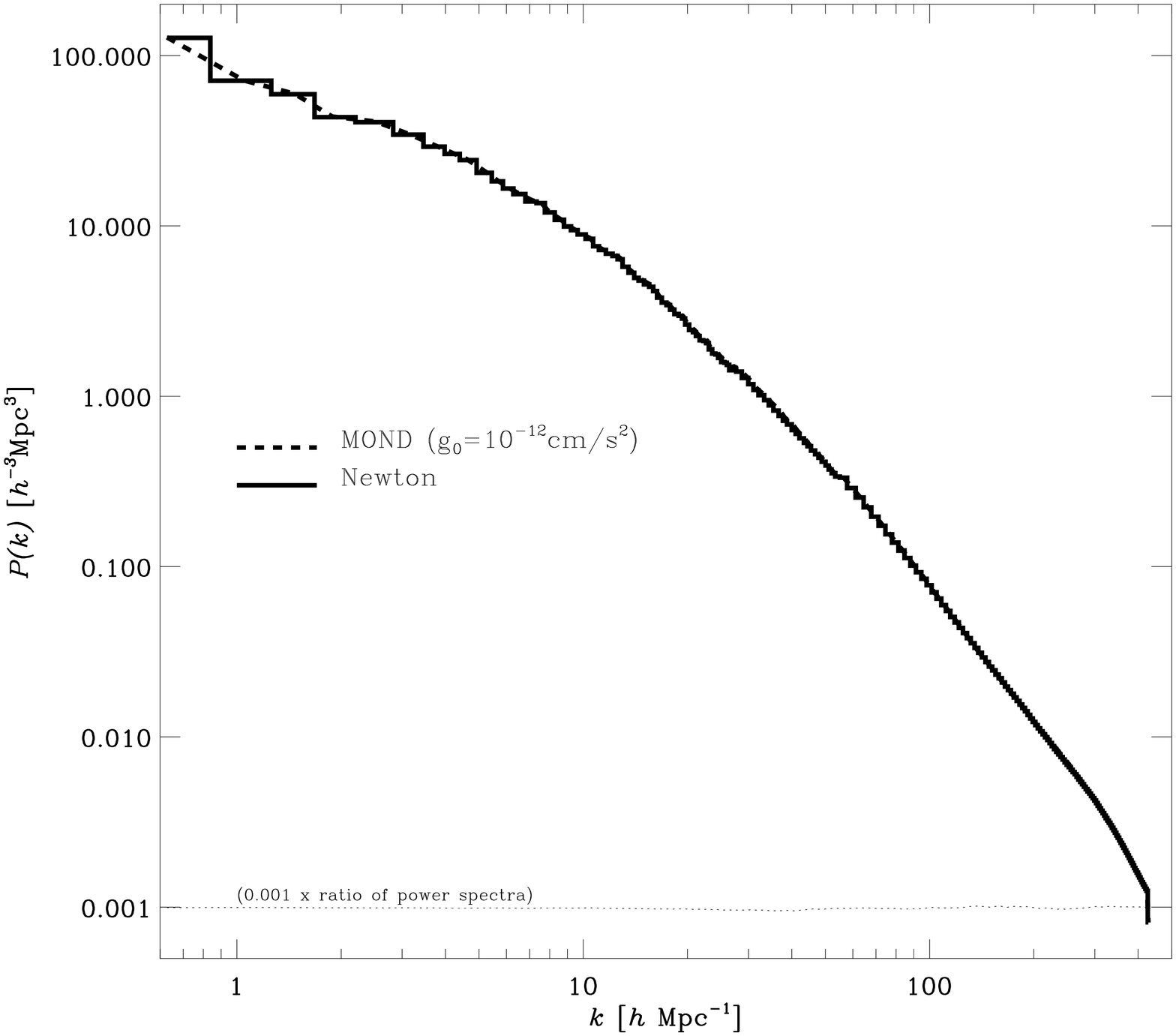}
    \caption{Power spectrum of the $64^3$ particle test simulation at
       redshift $z=0$.}
    \label{fig:power_064}
  \end{center}
\end{figure}

\begin{figure}
  \begin{center}
    \includegraphics[width=0.45\textwidth]{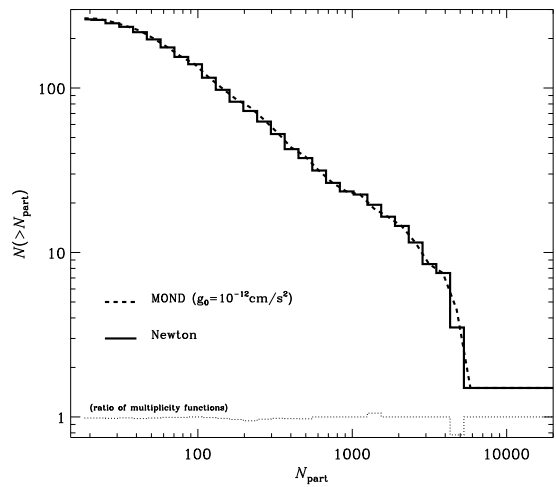}
    \caption{Mass function of the $64^3$ particle test simulation at
       redshift $z=0$.}
    \label{fig:massfunc_064}
  \end{center}
\end{figure}

Having established the credibility of our numerical integrator for
\Eq{eq:PoissonMOND}, we now test the temporal evolution of our code by
using a cosmological simulation in the Newtonian limit, i.e. lowering
the MONDian acceleration $g_0$ to such a low value that it will not
impact upon structure formation.  For that purpose we used
$g_0=10^{-12}$cm/s$^2$ and compare the final output to a simulation
run with the standard Newtonian \nbody\ code. The simulation
fascilitated $64^3$ particles in a box of comoving side length
$15$\hMpc. The cosmology is characterized by $\Omega_0$=0.3,
$\Omega_{\Lambda, 0}$=0.7, $\sigma_8$=0.9, and $h=0.7$ but actually of
no relevance for this particular test.

A first visual impression of the density field in given in
\Fig{fig:density_064} where we show the grey-scaled density at each
particle position of the simulation at redshift $z=0$. The differences
are at best marginal indicating that the modified solver performs
correctly in the Newtonian limit. A more quantitative comparison can
be found in \Fig{fig:power_064} where we show the matter power
spectrum at redshift $z=0$. The curves are indistinguishable which is
proven by the ratio of the two curves plotted as a dotted line (lowered
for clarity by the factor 0.001). We further ran the MPI enabled halo
finder \ahf\footnote{\ahf\ is the successor of \mhf\ introduced in
  \citet{Gill04a}.}  (Knollmann et al. 2008, in prep.)  over both
simulations. The resulting mass functions of identified objects is
given in \Fig{fig:massfunc_064}. Again, there are at best differences
that are readily ascribed to numerical errors during the time
integration due to our vastly different schemes of solving Poisson's
equation in the two test runs.

\section{Cosmological simulations} \label{sec:cosmology}

To date there is only a limited number of cosmological simulations
aiming at studying the effect sof MOND in a full cosmological context
\citep[][KG04]{Nusser02}. However, both these investigations
``simply'' solved for the Newtonian forces and modified them according
to a non-cosmological MOND prescription prior to being used with the
equations-of-motion (cf. \Eq{eq:eom}). They both showed that, under
their respective heuristic formulations, MOND can lead to large-scale
clustering patterns aking to a \LCDM\ model.

In the present study we are going to abandon (at least) one of the
assumptions made by these authors, namely that $\mathbf C=\mathbf 0$
and hence that \Eq{eq:curl} can be inverted to obtain a function
$\mathbf g_M(\mathbf g_N)$. Here we numerically integrate the MONDian
analogue to Poisson's equation~(\ref{eq:PoissonMOND}) which directly
leads to considering the influence of the yet unkown and unregarded
curl-field $\mathbf C$ in the process of structure formation. The
``only'' assumption we adhere to is that MOND does not affects
fluctuations and leaves the background cosmology intact \citep[e.g.,][]{Nusser02, Knebe04}. We defer to a later study that will make
use of modified Friedmann equations accounting for the effects of
MOND.

\subsection{The Simulation Details}

\begin{table}
  \caption{Model parameters. In all cases a value for the Hubble parameter
    of $h=0.7$ was employed. For model OCBMond we explicitly assumed $\mathbf C=\mathbf 0$ 
    whereas OCBMond2 is based upon a numerical integration of the MONDian Poisson's
    equation as described in \Sec{sec:code}.}
\label{tab:simulations}
\begin{tabular}{lllllll}

label   & $\Omega_0$ & $\Omega_b$ & $\lambda_0$ & $\sigma_8^{z=0}$ & $\sigma_8^{\rm norm}$ & $g_0$ [cm/s$^2$]\\ 
\hline \hline
\LCDM\   &    0.30    &   0.04    & 0.7  &   0.88    & 0.88  & ---\\
OCBMond  &    0.04    &   0.04    & 0.0  &   0.92    & 0.40  & 1.2 $\times 10^{-8}$ \\
OCBMond2 &    0.04    &   0.04    & 0.0  &   0.92    & 0.40  & 1.2 $\times 10^{-8}$ \\

\end{tabular}
\end{table}

Following KG04 and using the input power spectra derived with the
CMBFAST code \citep{Seljak96} we displace $128^3$ particles from their
initial positions on a regular lattice using the Zel'dovich
approximation \citep{Efstathiou85}. The box size was chosen to be
32\hMpc\ on a side. This choice guarantees proper treatment of the
fundamental mode which will still be in the linear regime at $z=0$
(cf. the scale turning non-linear at $z=0$ is roughly 20\hMpc\ for the
models under investigation). The particulars (and terminology) of the
models under investigation are summarized in \Tab{tab:simulations}. 

As we used the (Newtonian) Zeldovich approximation to generate the
initial conditions we need to make sure that the universe is still
Newtonian at the starting redshift. To this extent our choice for the
(free) function $\gamma(a)$ in \Eq{eq:PoissonMOND} is $\gamma(a) = a
g_0$. This ensures that $\mu(\mathbf x) = 1$ and hence
\Eq{eq:PoissonMOND} reduces to the Newtonian case
\Eq{eq:PoissonNewton}. We need to acknowledge that this differs from
the treatment in KG04 who actually used $\gamma(a)=g_0$; their
simulations did not start early enough and were in parts already
MONDian at the initial redshift. However, as we will see later on
during the analysis of the runs that our results hardly differ from
the findings of KG04. We further like to mention that the present
study does not aim at distinguishing one MOND theory from another!  We
merely apply our novel solver to one particular MOND model of our
choice showing that this implementation can lead to results that are
at least comparable to the favoured \LCDM\ structure formation
scenario. Besides, we also like to study the effects of the yet
unregarded and unkown curl-field $\mathbf C$ upon gravitational
structure formation, irrespective of the applicability of our
particular MOND model to reality; we leave a detailed study of various
MOND realisations to a future paper.

The particles were evolved from redshift $z=50$ until $z=0$ and in all
three runs a force resolution of 6\hkpc\ was reached in the highest
density regions. The mass resolution of the runs is $m_p = 1.30\times
10^{9}$\hMsun\ for the \LCDM\ model and $m_p = 0.17\times
10^{9}$\hMsun\ for the two low-$\Omega_0$ models, respectively. We
output 26 snapshots of the particle positions and velocities equally
spaced in time $t$ inbetween redshifts $z=5$ and $z=0$. These outputs
are then analysed with respect to their large-scale clustering
patterns as well as properties of individual objects employing the MPI
enabled halo finder \ahf\ again. For that purpose we though switched
off the unbinding procedure; we rather collect all particles within a
certain region of a local density peak and refer to them as ``object
particles'' defining the properties of that object. To be more
precise, \ahf\ locates density maxima in the simulation by invoking
the original adaptive-mesh hierarchy again used during the simulation
procedure. For each of these peaks we step out in logarithmically
spaced radial bins until the mean density inside that (spherical)
sphere drops below a fiducial value $\Delta \times \rho_b$ where
$\rho_b$ is the cosmological background density, i.e.

\be
 \frac{M(<R)}{\frac{4\pi}{3} R^3} = \Delta \times \rho_b ,
\ee

\noindent
defines the edge $R$ of an object. However, one needs to carefully
choose the correct overdensity $\Delta$ which is much higher for the
OCBMond models due to the low $\Omega_0$ value. The parameters used are
$\Delta=340$ for \LCDM\ and $\Delta=2200$ for OCBMond
\citep[see][Appendix C, and references therein]{Gross97}.
\footnote{Note that $\Delta$ depends both on redshift and cosmology.}

\subsection{The Matter Field} \label{sec:matterfield}

\begin{figure*}
  \begin{center}
    \includegraphics[width=0.95\textwidth]{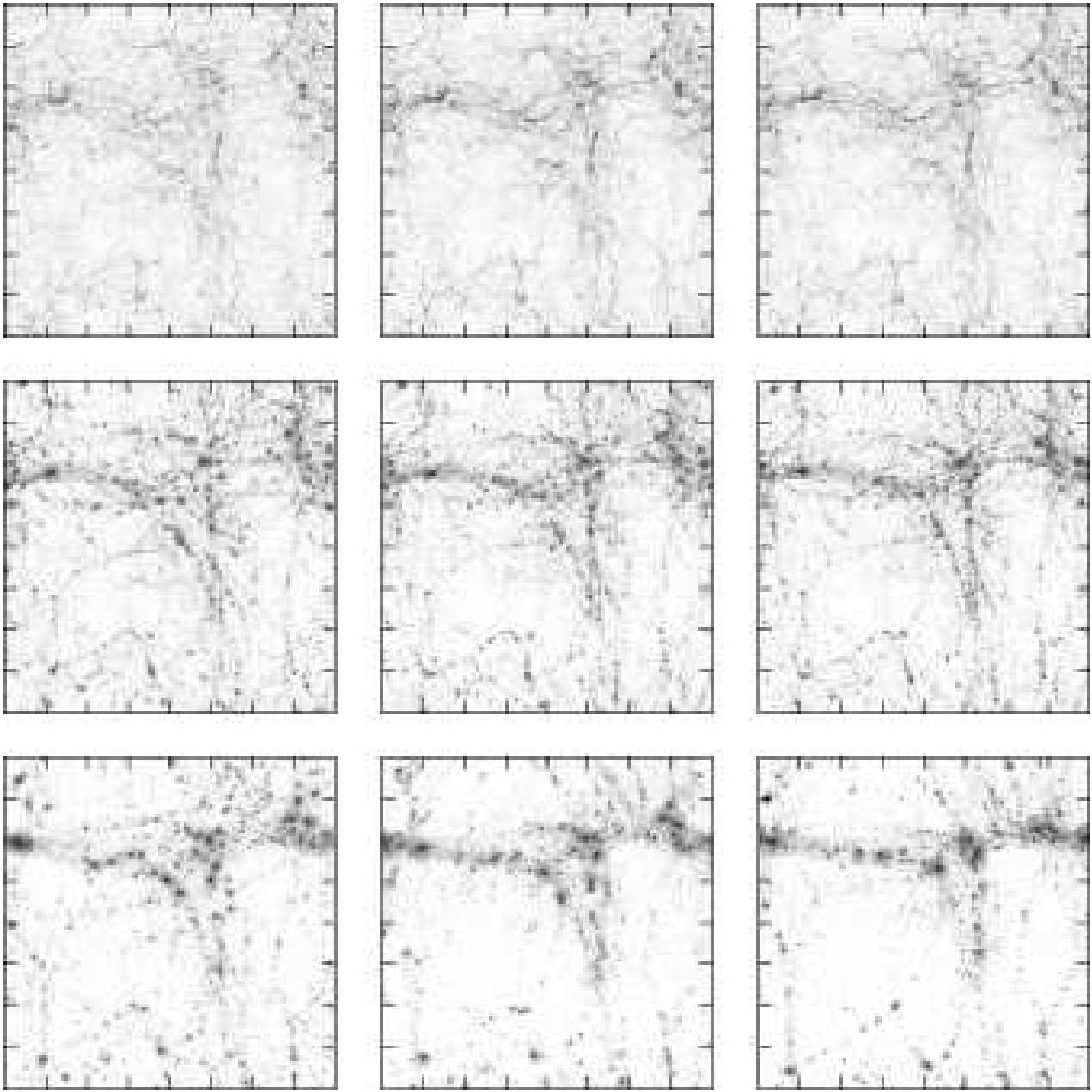}
    \caption{Density distribution at redshift $z=0$ (bottom row),
      $z=2$ (middle row), and $z=5$ (upper row) for the 128$^3$
      particle \LCDM\ (left column), OCBMond (middle column), and
      OCBMond2 (right column) simulations.}
    \label{fig:density_128}
  \end{center}
\end{figure*}

\begin{figure}
  \begin{center}
    \includegraphics[width=0.45\textwidth]{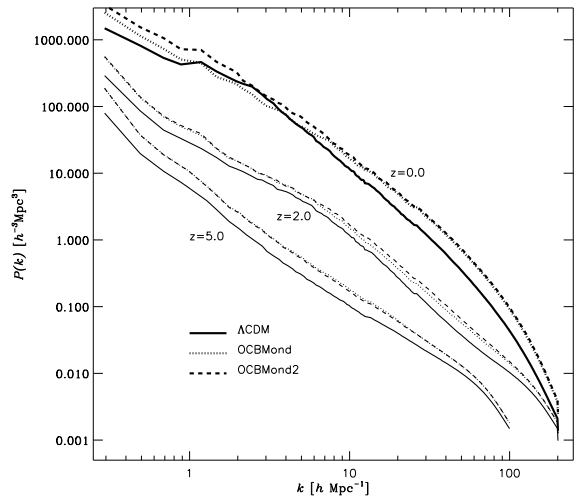}
    \caption{Dark matter power spectrum of all simulations at
      redshifts $z=5$, $z=2$, and $z=0$.}
    \label{fig:power_128}
  \end{center}
\end{figure}

\begin{figure}
\begin{center}
\includegraphics[width=.45\textwidth]{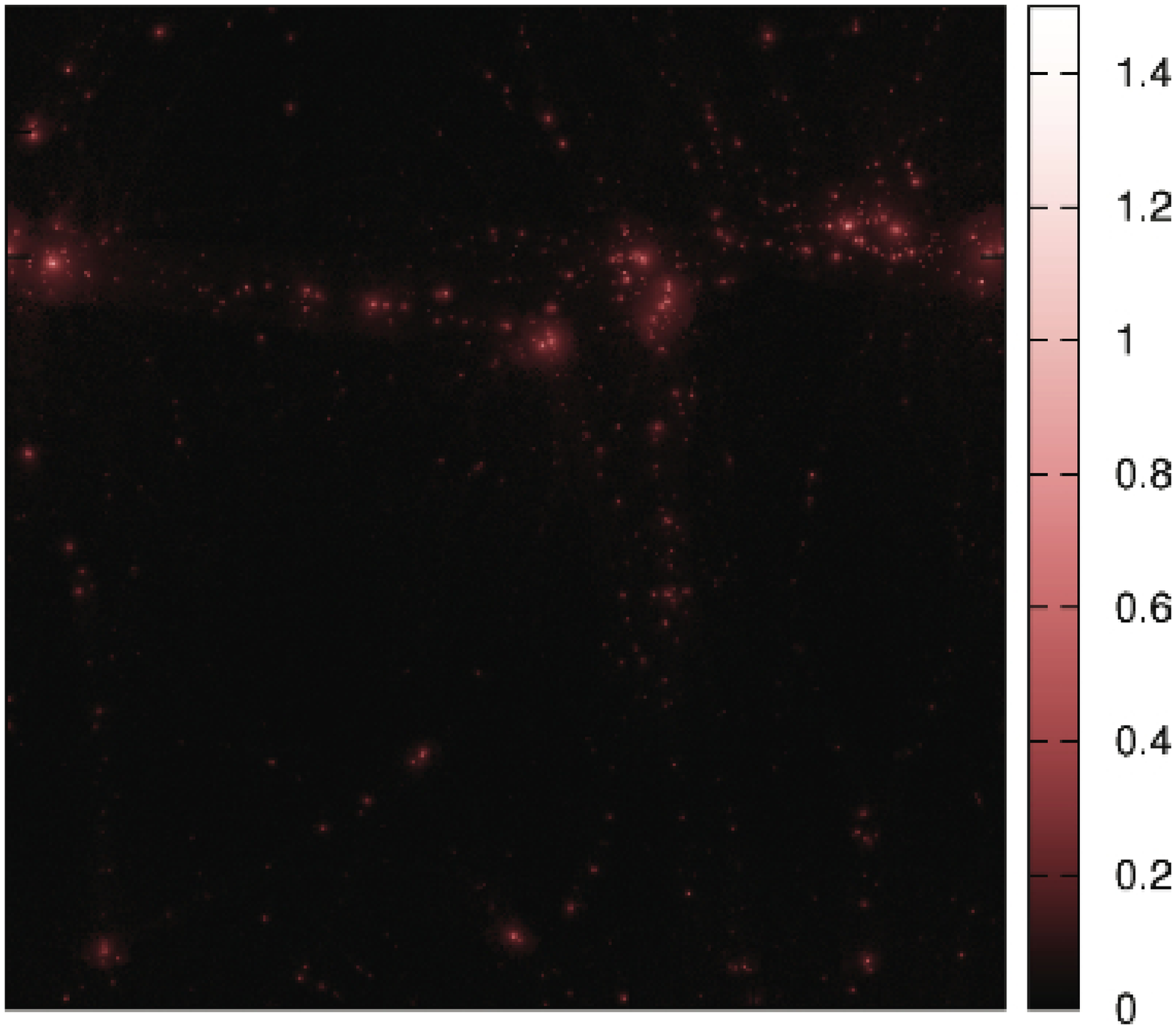}
\caption{Argument of the MONDian interpolation function $\mu(x)$ at
  redshift $z=0$ in the OCBMond2 model, i.e. the modulus of the
  MONDian force $|\nabla \Phi_M|$ normalized by $a^2g_0$. The values
  have been interpolated to the particle's positions and are
  colour-coded according to the scale shown on the right. For values
  above 1 the particles are treated according to Newtonian physics
  whereas values below 1 indicate the MONDian regime.}
\label{fig:mondian_for_in_particles}
\end{center}
\end{figure}

We start with inspecting the matter density field of all three runs in
\Fig{fig:density_128}. There we can see that at redshift $z=0$ the
differences amongst the models are rather small. However, there are
major differences at larger redhifts due to the fact that the
formation of structures under MOND kicks in later and then at an
increased growth rate (i.e. $\delta \propto a^2$ in MOND as opposed to
$\delta \propto a$ for Newtonian physics, e.g.  \citet{Sanders01,
  Nusser02, Knebe04, Sanders08}). But the more interesting observation
is the comparison between our two MONDian implementations. There we
can see some objects that are still approaching each other in OCBMond
to have already merged in OCBMond2 indicative of an advanced
evolutionary stage.

We quantify the evolutionary stage in \Fig{fig:power_128} where we
present the matter power spectra at redshift $z=0$, $z=2$, and
$z=5$. We basically recover the same features as reported by KG04
(even though they considered a rather different MOND model with
$\gamma(a)=g_0$), namely the lack of a distinctive ``break'' due to
the transfer of power from large to small scales in the MONDian runs
and the (marginally) larger amplitude of power on scales $k\leq
1h$Mpc$^{-1}$ close to the fundamental mode. However, we also observe
that OCBMond2 evolved marginally faster on large scale and at later
times than OCBMond.

A rather natural question arises about the sites where MOND is
actually affective. To shed some light on this issue we show in
\Fig{fig:mondian_for_in_particles} the modulus of the MONDian force
$\vecg_{M2}$ in the OCBMond2 model at the particle positions at
redshift $z=0$ normalized by $a^2 g_0$ (which corresponds to the
argument of the MONDian interpolation function used with
\Eq{eq:PoissonMOND}). We find that particles in low-density regions
and the vicinity of objects, respectively, are in the MOND regime
whereas particles residing at the centres of objects
(i.e. high-density regions) are mostly unaffected by MOND. This may
result in different (hierarchical?) structure formation scenarios as
matter infall onto objects will most certainly be affected by the
``MONDian environment'' of objects. We will investigate this in the
following Section.

\subsection{Hierarchical Structure Formation} \label{sec:hierarchicalstructureformation}
\begin{figure}
  \begin{center}
    \includegraphics[width=0.45\textwidth]{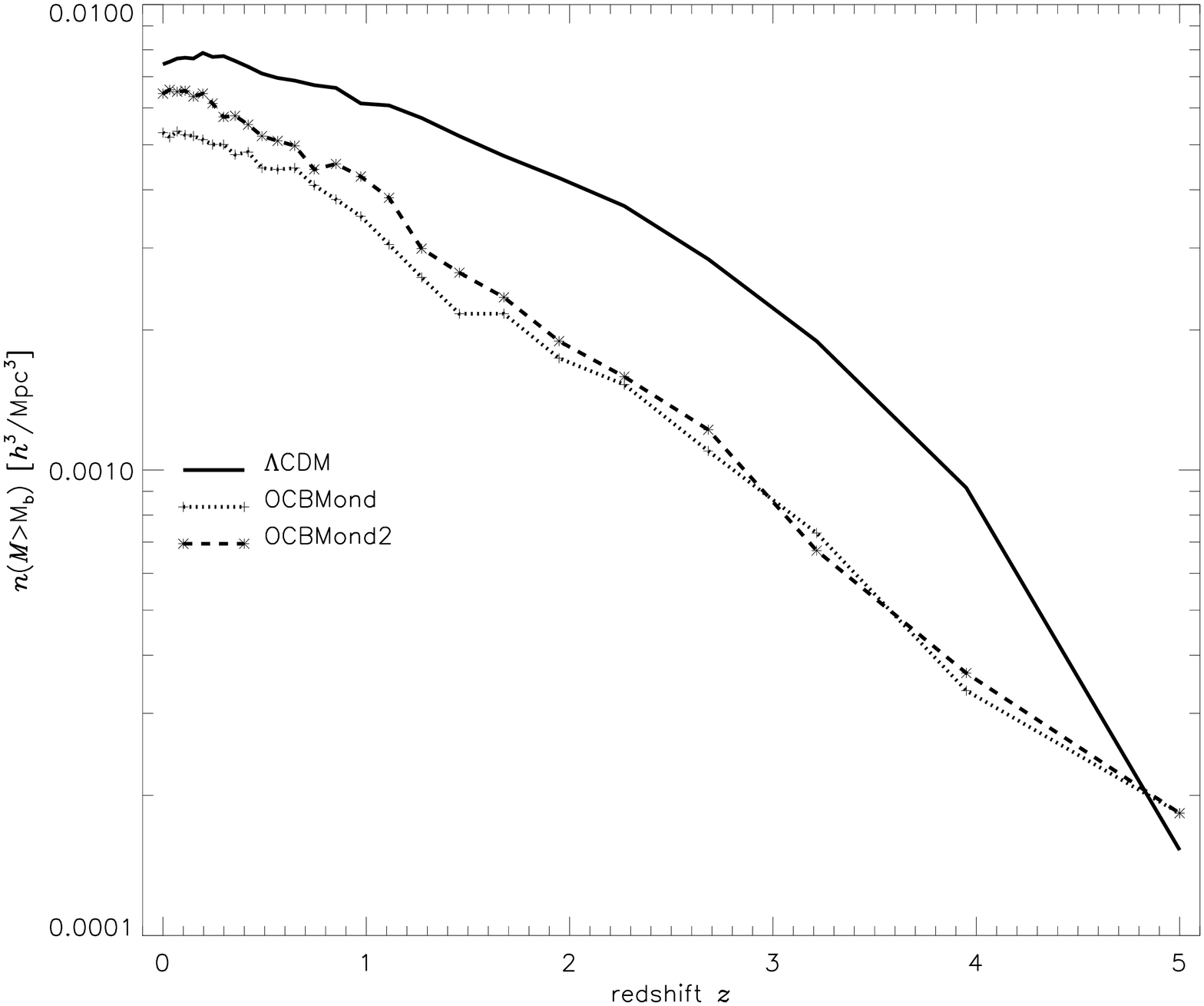}
    \caption{Redshift evolution of the abundance of objects with
      (baryonic) mass $M_b>10^{11}$\hMsun. Note that the baryonic mass
      for the \LCDM\ model has been estimated by multiplying the dark
      matter mass by the baryonic fraction.}
    \label{fig:abundance_128}
  \end{center}
\end{figure}

\begin{figure}
  \begin{center}
    \includegraphics[width=0.35\textwidth]{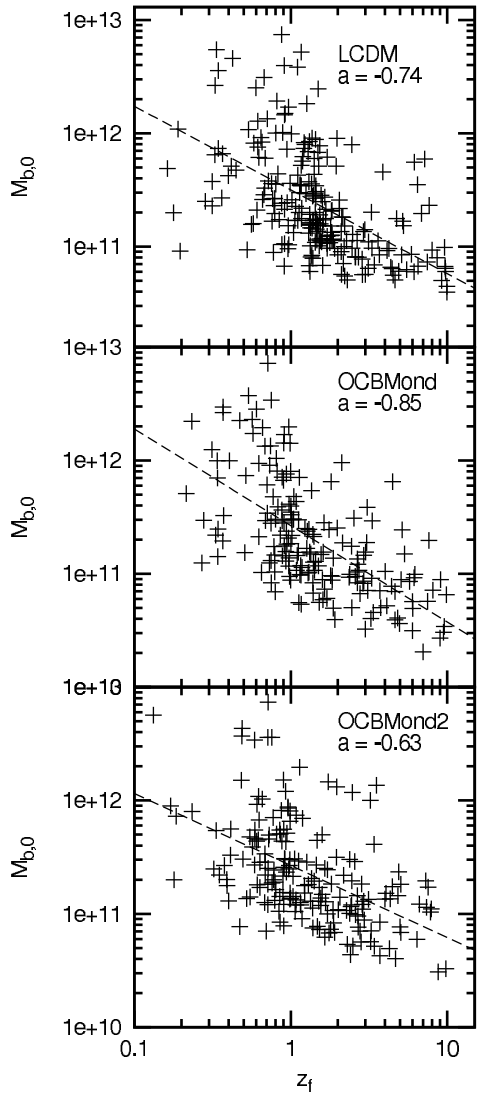}
    \caption{The ``baryonic'' masses $M_{b,0}$ of objects at today's
      redshift versus their formation redshift $z_f$. The solid lines
      are best-fit power laws to the scatter plot with the exponent of
      the power law given in the upper right corner of each panel.}
    \label{fig:M0_zf}
  \end{center}
\end{figure}

\begin{figure}
  \begin{center}
    \includegraphics[width=0.45\textwidth]{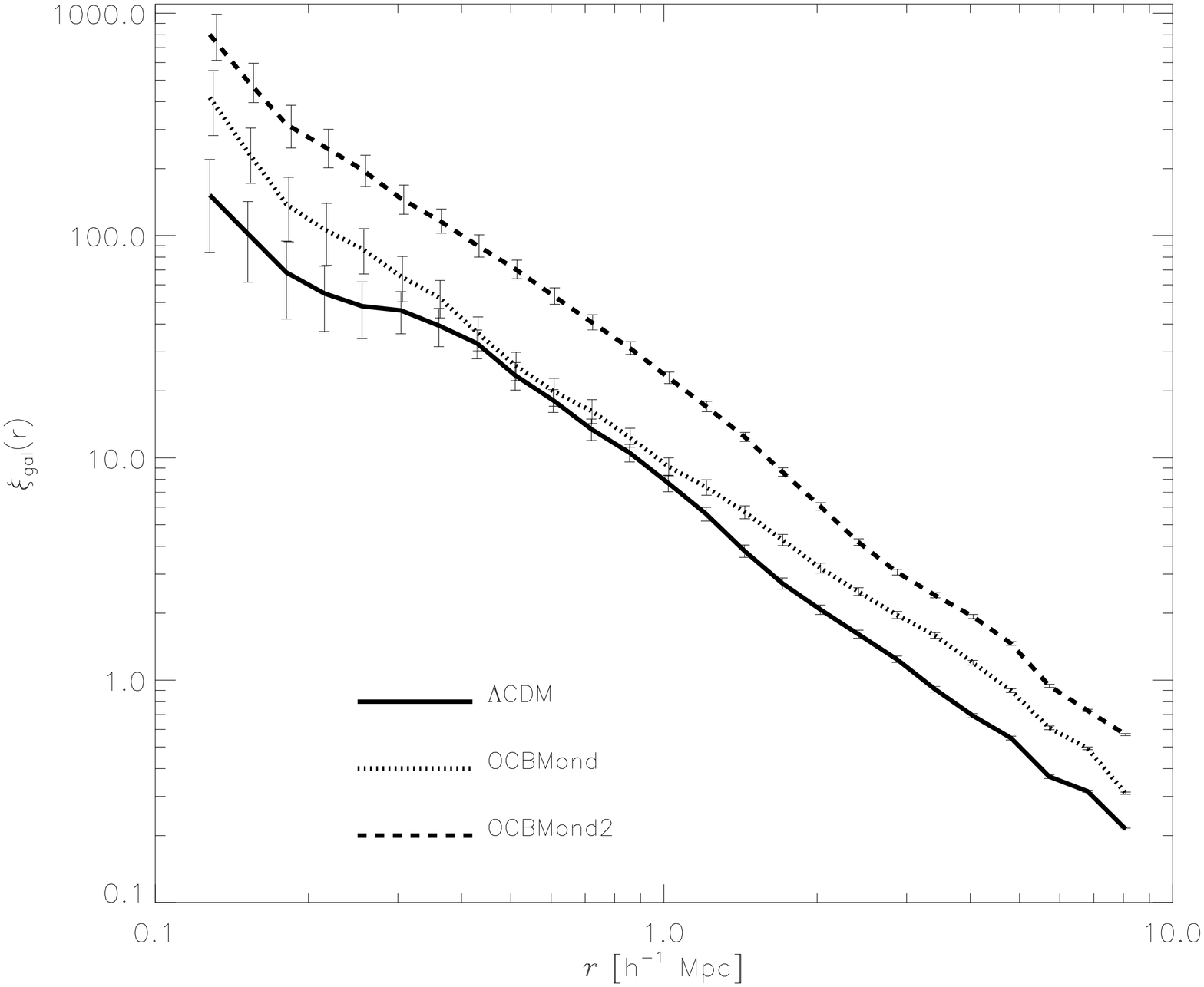}
    \caption{Two-point correlation function at redshift $z=0.0$ for
      the 1000 most massive objects. Error bars denote Poissonian
      error bars derived from the number of pairs in the respective
      distance bin.}
    \label{fig:xi_128}
  \end{center}
\end{figure}

\begin{figure}
  \begin{center}
    \includegraphics[width=0.45\textwidth]{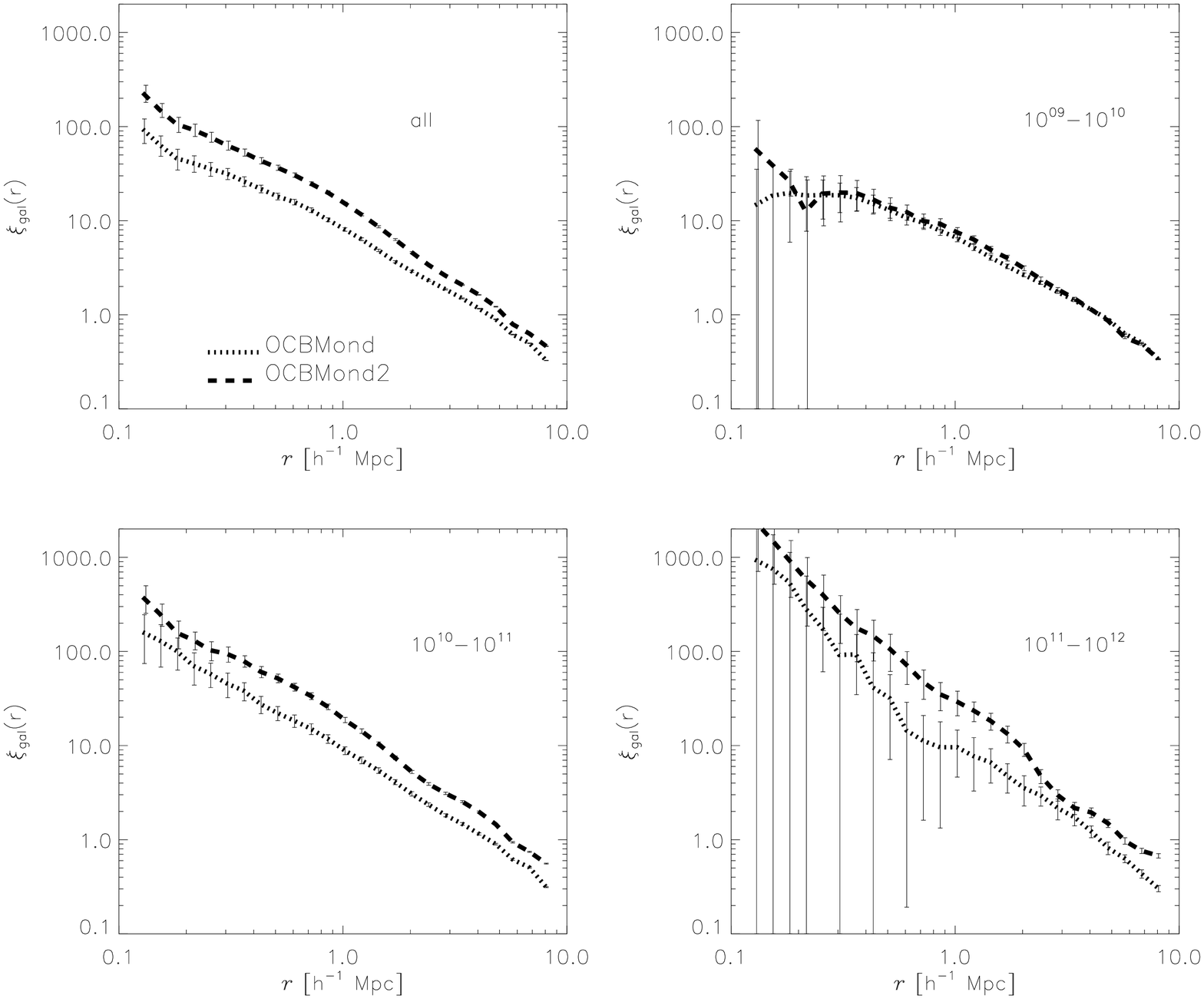}
    \caption{The two-point correlation function for objects of
      different mass in OCBMond and OCBMond2. The considered mass
      range in the calculation of $\xi_{\rm gal}$ is given in the
      upper right corner of each panel. The error bars are the
      Poissonian errors again.}
    \label{fig:xi_128_MOND12}
  \end{center}
\end{figure}

As already noted in \Fig{fig:density_128} and \ref{fig:power_128}
there appears to be a marginally faster evolution of (large-scale)
structures in OCBMond2. This is further confirmed in
\Fig{fig:abundance_128} where we show the abundance evolution of
objects with mass $M>10^{11}$\hMsun. At around redshift $z=1.5$
OCBMond2 starts to develop more objects. Anyways, the late onset and
faster evolution of structure formation under MOND in general was
already reported by \citet{Sanders01}, \citet{Nusser02} and
KG04. However, here we not only used a differnet MOND model but also
revised the way of presenting the comparison between the two MONDian
and the Newtonian runs. As OCBMond and OCBMond2 are simulations of the
gravitational interactions of baryons whereas the \LCDM\ model
contains both baryonic and dark matter a fair comparision should
correct for that. Or in other words, an object of mass $M$ in one of
the MOND runs merely contains baryons whereas the corresponding object
in \LCDM\ contains both baryons and dark matter. As the idea of MOND
is to replace dark matter by a modification to Newtonian dynamics we
should ``remove'' the dark halo from the \LCDM\ object when performing
a cross-comparison. We therefore multiply the \LCDM\ masses by the
cosmic baryon fraction $f_b=\Omega_b/\Omega_0$ and refer to all masses
as ``baryonic mass'' $M_b$ even though the interactions of baryons
other than gravity are \textit{not} modelled in the simulations
presented in this study. We note that this is not necessarily correct
as galaxies probably contain fewer baryons than given by the cosmic
baryon fraction $f_b$ \citep[e.g., ][]{Gottloeber07, Okamoto08}. Our
``baryonic'' \LCDM\ masses are therefore considered an upper limit and
are likely to be smaller.

By revising the mass of \LCDM\ dark matter halos we note that the
abundance evolution is similar yet different when comparing \LCDM\
against the two MONDian models. Similar in a sense that the in KG04
advocated dramatic difference at high redshift is not as pronounced
anymore. However, the actual shape of the Newtonian and MONDian curves
are though rather different with a more gentle increase in the number
of objects in \LCDM\ (at least at redshifts smaller than $z\approx
2.5$).

To gain a better understanding of structure formation under MOND we
constructed a merger tree for each individual object identified at
redshift $z=0$. We then fitted the universal mass accretion history
formula to the data \citep{Wechsler02}

\be \label{eq:MAH}
 M(z) = M_{(z=0)} e^{-\alpha z} \ .
\ee

\noindent
Applying the half-mass criterion \citep[e.g.,][]{Tormen97} to the fits
for obtaining the formation redshift (i.e. the formation redshift is
implicitly defined via $0.5 = e^{-\alpha z_f}$)\footnote{Even though
  we stored 26 snapshots between redshifts $z=0$ and $z=5$ for each
  model we prefer to use the fitted mass accretion history as it will
  provide us with a more precise estimate for the formation time.} we
are able to quantify hierarchical structure formation. We expect small
objects to form first and then successively merge to form larger and
larger entities \citep[e.g.,][]{Davis85}. This can be verfied in
\Fig{fig:M0_zf} where we plot today's mass against the formation
redshift of that object. We note a clear hierarchical tendency even in
both MOND models. To quantify differences we fitted a simple power law
to the data

\be
M_{b,0} \propto z_f^a
\ee

\noindent
where the best fit $a$ is given in the upper right corner of the
respective panel. We find that especially in OCBMond2 high-mass
objects appear to form at earlier times than in \LCDM. Note that this
does not contradict the dearth of massive objects at high redshifts as
observed in \Fig{fig:abundance_128}. Here we are tracing back
individual objects whereas in \Fig{fig:abundance_128} we considered
all objects above a given mass at a certain redshift; \Fig{fig:M0_zf}
simply indicates that (some of) the high mass objects found at
redshift $z=0$ must have already been in place at a higher redshift
than in, for instance, \LCDM. Surprisingly there is also a difference
between OCBMond and OCBmond2 with the former being closer to
\LCDM. This highlights the relevance of the curl-field $\nabla\times
\vech$ introduced in \Eq{eq:curl} for structure formation.

One of the findings of the study by KG04 was that the (formation)
sites of MONDian objects are marginally more correlated. Can we
vindicate that the two-point correlation function $\xi_{\rm gal}$ of
MONDian objects (i.e. galaxies and hence the subscript ${\rm gal}$)
has a larger amplitude (at least) on small scales. This can be
verified in \Fig{fig:xi_128} where we plot $\xi_{\rm gal}(r)$ for the
1000 most massive objects in all three models. We confirm the finding
of KG04 that OCBMond has a (marginally) larger amplitude than \LCDM,
especially on small scales. We further acknowledge that OCBMond2 has a
substantially larger amplitude! In order to better gauge the orgin of
the difference between OCBMond and OCBMond2 we split our objects up
into different mass bins (i.e. $M_{\rm bin}^1 =
[10^{9}-10^{10}]$\hMsun, $M_{\rm bin}^2 = [10^{10}-10^{11}]$\hMsun,
$M_{\rm bin}^3 = [10^{11}-10^{12}]$\hMsun) and calculate $\xi_{\rm
  gal}$ for each mass bin independently. The results can be viewed in
\Fig{fig:xi_128_MOND12}. We observe that the differences are due to
the biased formation sites of high-mass objects. We like to mention
that the number of objects in each mass bin is practically identical
amongst OCBMond and OCBMond2; only in the range
$10^{11}-10^{12}$\hMsun\ there are about 24\% more objects in OCBMond2
as indicated by \Fig{fig:abundance_128}. We hence ascribe the bias
between OCBMond and OCBMond2 as found in \Fig{fig:xi_128} to the large
mass objects and hence the stronger evolution of large scales as
already seen in \Fig{fig:density_128} and~\ref{fig:power_128}.

\subsection{Cross-Correlation of Properties} \label{sec:crosscorrelation}
\begin{figure}
  \begin{center}
    \includegraphics[width=0.45\textwidth]{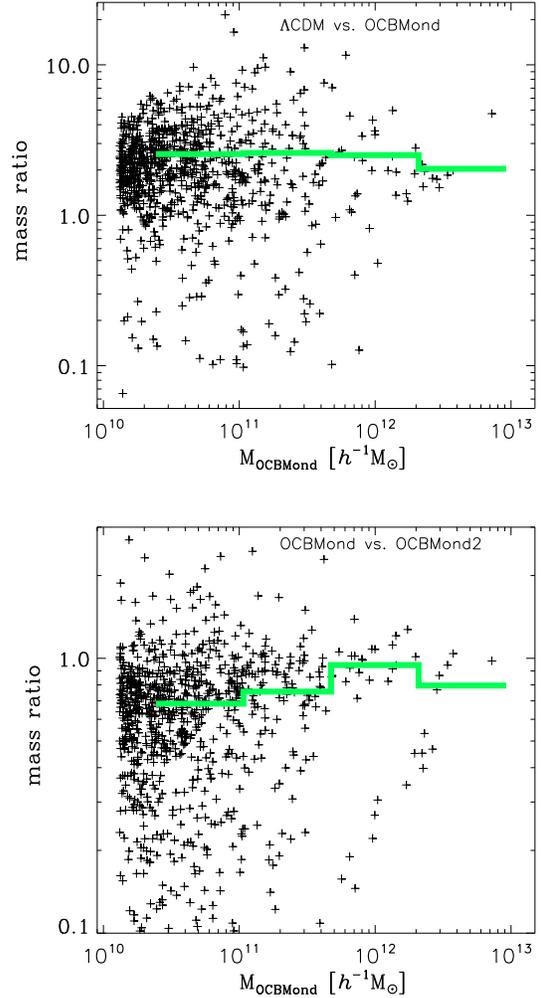}
    \caption{Ratio of masses for cross-identified objects at redshift
      $z=0$. The histograms represent the mean ratio in the respective
      bin. Note that the \LCDM\ masses have been lowered by the
      baryonic fraction to be comparable to the MONDian values.}
    \label{fig:MassMass_128}
  \end{center}
\end{figure}

\begin{figure}
  \begin{center}
    \includegraphics[width=0.45\textwidth]{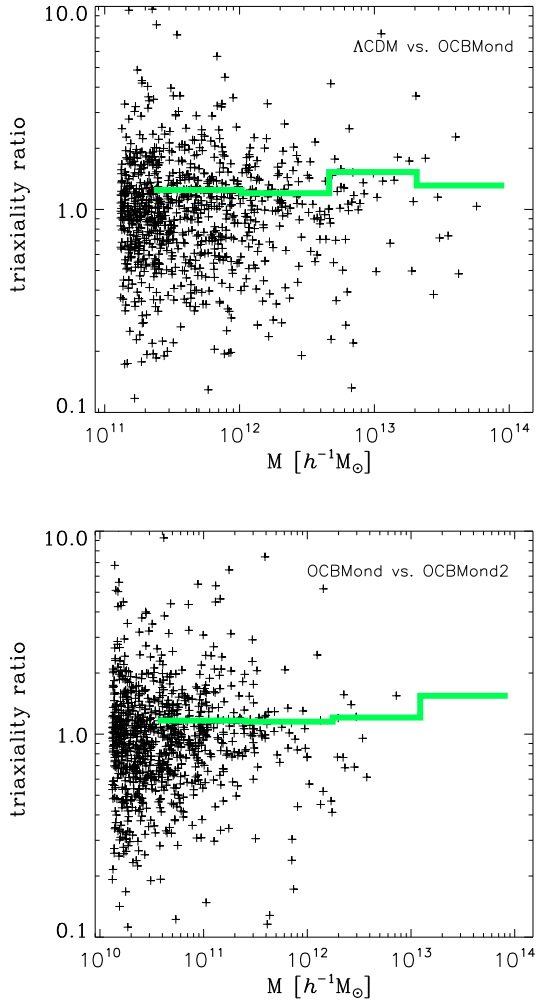}
    \caption{Ratio of triaxialities for cross-identified objects at redshift $z=0$. The
      histograms represent the mean ratio in the respective bin.}
    \label{fig:TT_128}
  \end{center}
\end{figure}

Something not considered in the study of KG04 is the cross-correlation
of individual objects. To this extent we use the tool
\texttt{MergerTree} that comes with the distribution of the \nbody\
code \amiga\ and serves the purpose of identifying corresponding
objects either in the same simulation at different redshifts (and
hence the name ``MergerTree'') or in simulations of different
cosmological models but run with the same initial phases for the
initial conditions. The cross-correlation is done by linking objects
that share the most common particles.

We start with the most simple yet still interesting quantify for our
cross-comparison, namely the mass $M$. In \Fig{fig:MassMass_128} we
show how $M_b$ correlates across \LCDM\ and OCBMond (upper panel) as
well as across OCBMond and OCBMond2 (lower panel). Remember that we
defined $M_b$ as the ``baryonic'' mass of our halos that coincides
with the actual total mass in the two MOND models but corresponds to
$M_{\rm tot} \times f_b$ for \LCDM; by the usage of the cosmic baryon
fraction $f_b = \Omega_b/\Omega_0$ in this formula we will have an
upper limit for the ``baryonic'' \LCDM\ mass
\citep[cf. ][]{Gottloeber07, Okamoto08}. Despite the ``baryonic''
correction we still observe a difference between the masses of
``identical'' objects in \LCDM\ and the MONDian runs. This can in
parts be ascribed to the use of the cosmic baryon fraction and in
parts to the definition of the edge of objects: remember that we use
$\Delta=340$ for \LCDM\ as opposed to $\Delta=2200$ for MOND. The
latter value translates into a smaller radius and hence less
mass. When applying the same density threshold of $\Delta=340$ to the
MONDian runs (not presented here though) the MONDian masses increase
by approximately 30\% bringing them into better (yet not perfect)
agreement with the \LCDM\ masses.

The other interesting observation in \Fig{fig:MassMass_128} is the
fact that there is a systematical difference between the masses of
cross-identified objects in OCBMond and OCBMond2. While the most
massive objects appear to have identical masses, there is a clear
trend for low-mass OCBMond2 objects to be more massive than their
OCBMond counterparts. This difference flattens to values closer to
unity when considering objects at higher redshift. This phenomenon can
therefore be ascribed and explained by the even stronger evolution of
the OCBMond2 simulation as already noticed in \Fig{fig:density_128}
and \Fig{fig:abundance_128}, respectively. As pointed out by
\Fig{fig:mondian_for_in_particles}, MOND is most affective in the
outer regions of objects and hence leaving its imprint via differing
infall patterns of material. We therefore ascribed the differences to
variation in the mass accretion histories as confirmed by
\Fig{fig:M0_zf}.

We close this Section with a brief investigation of the shape of
objects as defined by, for instance, the triaxiality parameter
\citep{Franx91}

\be \label{eq:triaxiality}
 T = \frac{a^2-b^2}{a^2-c^2}
\ee

\noindent
where $a>b>c$ are the principal axes of the moment of intertia tensor
we note that there are hardly any differences at all. This can be
verified in \Fig{fig:TT_128} where we plot for the same set of
cross-identified halos as already used in \Fig{fig:MassMass_128} the
ratio of the respective triaxialities as a function of mass
again. This figure highlights that despite variations in the mass
growth of objects at least their triaxialities are unaffected.

\subsection{The curl field} \label{sec:curlfield}

\begin{figure}
  \begin{center}
    \includegraphics[width=.45\textwidth]{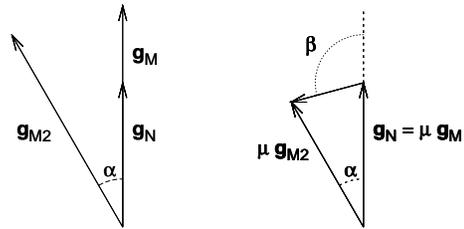}
    \caption{Sketch illustrating the vectors of \Eq{eq:curl_field} and
      the angles between them. $\vecg_N$ is the Newtonian force
      vector, $\vecg_M$ the one used with the OCBMond model, and
      $\vecg_{M2}$ the solution of our new solver responsbile to the
      evolution of the OCBMond2 model.}
    \label{fig:vectors_figure}
  \end{center}
\end{figure}

\begin{figure}
  \begin{center}
    \includegraphics[width=.45\textwidth]{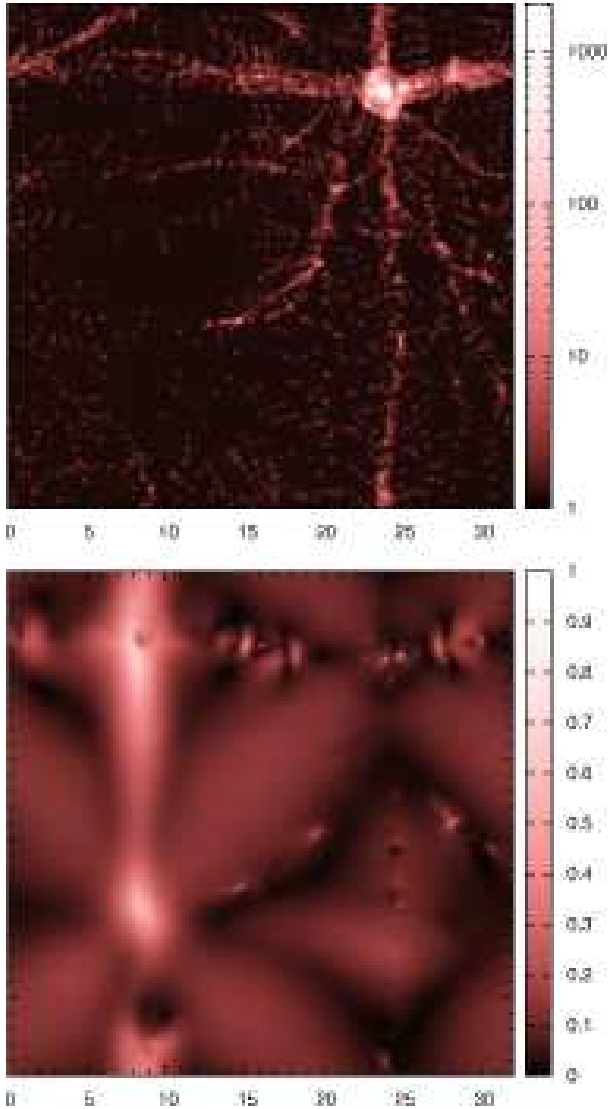}
    \caption{Slice of thickness 125\hkpc\ of the OCBMond2 simulation
      at redshift $z=0$. The slice is chosen to contain a strong
      density peak (i.e. it is centred about the position of the most
      massive object found in th simulation).  The upper panel shows
      the density as obtained on a regular 256$^3$ grid. The same grid
      is used to calculate the modulus of the curl field normalized by
      the modulus of the Newtonian force, i.e. $|\nabla \times
      \vech|/|\vecg_N|$, which is presented in the lower panel.}
    \label{fig:curl_maps}
  \end{center}
\end{figure}

\begin{figure}
  \begin{center}
    \includegraphics[width=.45\textwidth]{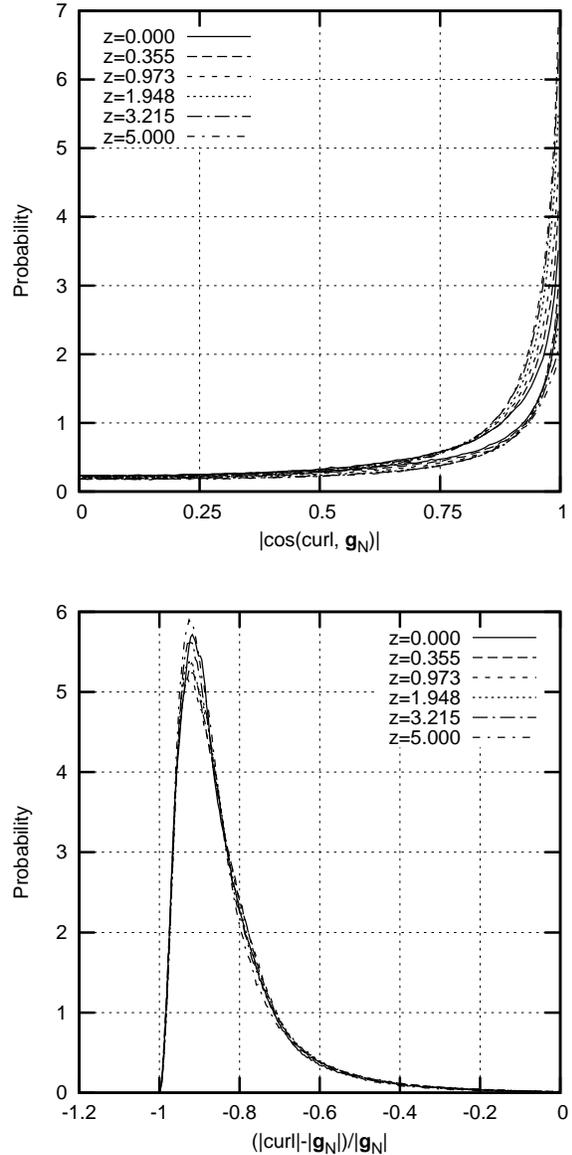}
    \caption{Probability distribution of the angle between the
      curl-field $\nabla\times\vech$ and the Newtonian force $\vecg_N$
      (upper panel) and the fractional difference between these two
      values (lower panel) at various redshifts.}
    \label{fig:curl_vs_newt}
  \end{center}
\end{figure}

\begin{figure}
  \begin{center}
    \includegraphics[width=.45\textwidth]{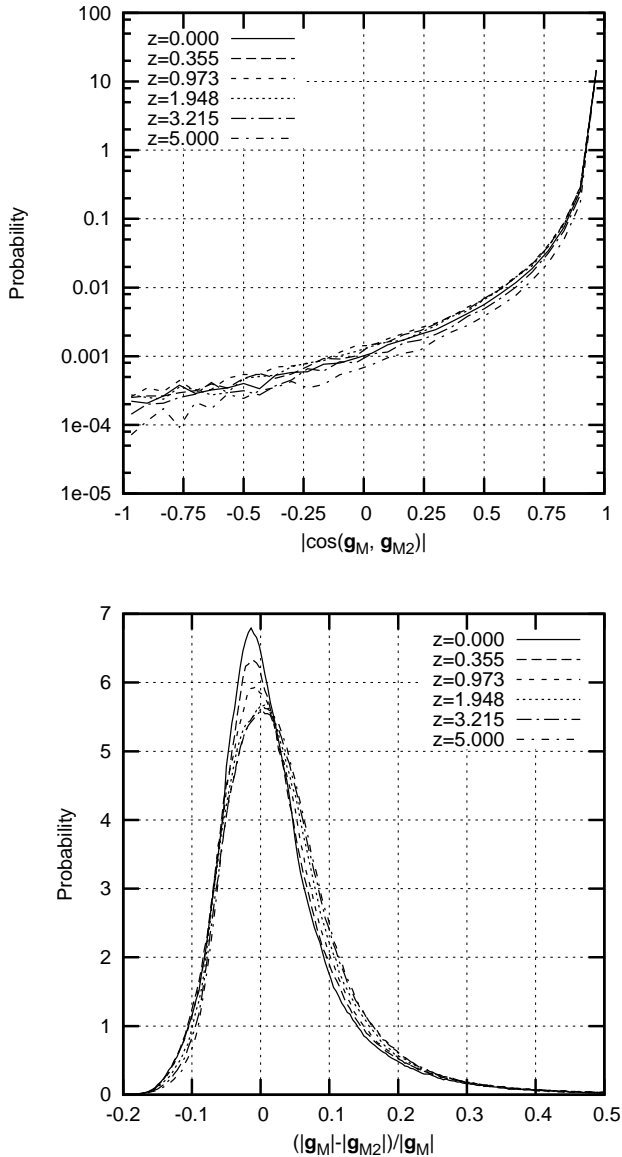}
    \caption{Probability distribution of the angle between the MONDian
      forces $\vecg_M$ and $\vecg_{M2}$ (upper panel) and the
      fractional difference between these two values (lower panel) at
      various redshifts.}
    \label{fig:mond_vs_mond2}
  \end{center}
\end{figure}

\begin{figure}
  \begin{center}
    \includegraphics[width=.5\textwidth]{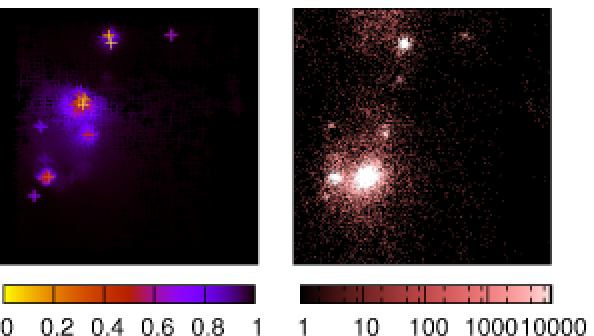}
    \caption{Projection of a sub-box of side length $1.5$\hMpc\ at
      redshift $z=0$. The left plot shows a colour-representation of
      $\cos{(\vecg_M,\vecg_{M2})}$ whereas the right plot shows the
      density field. Both values are evaluated at the particles'
      positions. Note that both figures are centred on the same
      point.}
    \label{fig:crazy_forces}
  \end{center}
\end{figure}

As we have just seen there are subtle yet noticable differences when
using our novel Poisson solver for the MONDian
equation~(\ref{eq:PoissonMOND}) as oppossed to the simplified
prescription of inverting \Eq{eq:curl} under the assumption of
$\mathbf C=\mathbf 0$. In that regards it appears mandatory to study
the origin of these devations. We therefore present in this Section a
detailed investigation of the curl-field $\nabla \times \vech$ and its
effects upon structure formation.

Given the solution of the \Eq{eq:PoissonNewton} and \Eq{eq:PoissonMOND} the
curl-field is readily calculated

\be \label{eq:curl_field}
\nabla\times\textbf{h} = 
\left[ \mu\left(\frac{\nabla\Phi_M}{a^2 g_0}\right)\nabla\Phi_M\right] - 
\nabla\Phi_N.
\ee

\noindent
Note that setting $\nabla\times\vech=0$ corresponds to the OCBMond
model.

In order to clarify all definitions and terminology to be used from
now on we show in \Fig{fig:vectors_figure} a sketch of all the vectors
$\vecg_N=-\nabla \Phi_N$, $\vecg_M=-\nabla\Phi_M$, and
$\vecg_{M2}=-\nabla\Phi_{M2}$ in play. The assumptions used with the
OCBMond model imply that the forces there and the Newtonian forces,
$\vecg_M$ and $\vecg_N$, are parallel whereas they both can have an
angle $\alpha$ with the OCBMond2 force vector $\vecg_{M2}$. This angle
is due to the influence of the curl-field and can be described by the
angle $\beta$ (cf. \Eq{eq:curl_field}).

Interpreting the curl-field as a modification to $\mathbf g_N$ on the
right-hand side of \Eq{eq:curl} the question arises about the actual
adjustment to (the implicitly defined) $\vecg_{M2}$ stemming from this
additional "source" term. We therefore normalize $\vecC$
by the Newtonian acceleration $\vecg_N$ and study its spatial
dependence. This can be viewed in \Fig{fig:curl_maps} where we show
$|\vecC|/|\vecg_N|$ (lower panel) in a slice of
thickness 125\hkpc\ in the OCBMond2 simulation as calculated on a
256$^3$ grid.  We do observe a marginal correlation of the
``correction to the source'' with the actual matter density shown in
the upper panel. We further note that the modulus of the curl-field is
actually always smaller than the Newtonian force, however, its spatial
variation is though highly non-trivial. In some (high-density) regions
there appear to be some kind of ``shocks'' where the curl-field
changes dramatically changes its value. The only conclusion we can
draw from this qualitative analysis is that the curl-field has an
influence primarily in low-density regions: its value is close to the
Newtonian force in the void regions. Finally we like to remark on the
``diamond shapes'' seen in \Fig{fig:curl_maps}; they are an artifact
of the periodic boundary conditions and stem from the periodic images
of the largest object/void. These attracting images create the
observed symmetry leading to the diamond shapes.

In order to make a more quantitative analysis of the curl-field, we
use the snapshots of the simulation OCBMond2 and calculate the
Newtonian as well as the two MONDian forces at various
redshifts. These forces are then extrapolated to the particle
positions in the same way as done during the time integration of
\Eq{eq:eom}. We are aware that this extrapolation will bias our
results towards high-density regions (where the particles actually
reside) and ignore the effects-of-interest in low-density regions,
respectively. However, our integration scheme is based upon the idea
of sampling phase-space with particles and hence the \textit{forces at
  the particle positions} are relevant for the time integration of
\Eq{eq:eom} and hence the evolution of the simulation.

In \Fig{fig:curl_vs_newt} we now present a quantitative comparison of
the curl field, ${\rm curl}=\nabla\times\vech$, and the Newtonian
forces $\vecg_N$ by plotting the probability distribution of the angle
between the two vectors (upper panel) as well as the relative
difference of the modulus (lower panel). First, we note that there is
practically no evolution with redshift. Second, the curl-field is well
aligned with the Newtonian force though it may also point in the
opposite direction! In order to emphasize the skewness of the
distribution about $\cos{({\rm curl}, \vecg_n)} = 0$ we show the
modulus of it. The upper curves thereby correspond to $\cos{({\rm
    curl}, \vecg_n)} < 0$, i.e. it is more likely that the curl points
in the opposite direction. However, the lower panel of
\Fig{fig:curl_vs_newt} indicates that the actual change to the
``source'' term for the implicitly defined $\vecg_{M2}$
(cf. \Eq{eq:curl}) induced by the curl is rather small. Here we
show the fractional difference between the moduli of the curl and the
Newtonian force vector. This distribution peaks at approximately
$-0.9$ and hence a $\approx$10\% correction to $\vecg_N$ in
\Eq{eq:curl}; this modification is again independent of
redshift. So, the net effect of the curl field is to reduce the
``source'' on the right hand side of \Eq{eq:curl}. We expect this
to translate into a reduction of $\vecg_{M2}$ with respect to
$\vecg_M$, too.

But because of the implicit definition and the vector-nature of the
quantities involved in the calculation of $\vecg_{M2}$ it is difficult
to make predictions for the change in $\vecg_{M2}$ induced by the
modifications to the ``source'' in \Eq{eq:curl}. We therefore plot
in \Fig{fig:mond_vs_mond2} the analogous quantities as in
\Fig{fig:curl_vs_newt} this time for the comparison of $\vecg_M$ and
$\vecg_{M2}$; this should reveal the effects of the curl-field
directly. As the angle distribution is no longer symmetric we expand
it over the whole range from $\cos{(\vecg_M,\vecg_{M2})} \in
[-1,+1]$. The upper panel (showing the angle distribution) clearly
indicates that both MONDian forces are \textit{well} aligned (note the
logarithmic scale on the $y$-axis). The lower panel (showing the
fractional difference between $\vecg_M$ and $\vecg_{M2}$) proofs what
we already expected: the distribution is centered about $0$ but shows
a skewness towards values of stronger $\vecg_M$ and lower
$\vecg_{M2}$, respectively. This skewness is a manifestation of the
result obsvered in \Fig{fig:curl_vs_newt}, namely that the curl
preferentially decreases the ``source'' in \Eq{eq:curl}.

We further note in \Fig{fig:mond_vs_mond2} that there is a marginal
redshift evolution of the peak in the distribution of the fractional
differences of the two MONDian forces. The peak gradually migrates
away from $0$ at redshift $z=5$ towards more negative values of
approximately $-0.09$ at $z=0$. While it appears counter-intuative to
explain the advanced structure formation in OCBMond2 found in previous
Sections (cf. \Fig{fig:density_128}, \Fig{fig:power_128},
\Fig{fig:abundance_128}, etc.) with the finding that the forces are
smaller in that model, we rather believe that the shift in the peak
can be held responsible for it. This shift towards negative values
translates into \textit{stronger OCBMond2} forces and hence structure
formation at an accelerated speed.

As an illustrative example of the misalignment between the two MONDian
forces (and its relation to the underlying density field) we show in
\Fig{fig:crazy_forces} both the (colour-coded) cosine of the angle
between $\vecg_M$ and $\vecg_{M2}$ (left panel) alongside the density
at the particles' positions at redshift $z=0$. We note that there
is no apparant correlation of the misalignment with density.

\section{Summary \& Conclusions} \label{sec:conclusions}

We presented a novel solver for the analogue to Poisson's equation
taking into account the effects of modified Newtonian dynamics (MOND).
This equation is a highly non-linear partial differential equation for
which standard solvers based upon Fourier transformation techniques
and/or tree structures are not applicable anymore; one has to defer to
multi-grid relaxation techniques \citep[e.g.,][]{Brandt77, Press92,
  Knebe01} in order to numerically solve it.

The major part of this paper hence deals with the necessary
(non-trivial) adaptions to the existing multi-grid relaxation solver
\amiga\ (successor to \mlapm\ introduced by \citet{Knebe01}). We show
that the accuracy of our MOND solver is at a credible level for a
static problem with known MONDian solution and that it reproduces
correct results in the Newtonian limit for cosmological simulations.

In the case that we don't want to adhere to any (undjustifyable and
hence ad-hoc) assumptions as, for instance, done by KG04, we are
facing the problem that there is still a lot of freedom from the
covariant point of view to describe the original phenomenological MOND
theory.  To be able to actually perform MONDian cosmological
simulations we decided to choose one particular model for MOND and to
leave for later studies the analysis of the effects that differents
theories producces on the structure formation.  One of the
suppositions KG04 made relates to the relation between the MONDian and
the Newtonian force vector. They use the rather simple relation as
given by setting $\mathbf C=\mathbf 0$ in \Eq{eq:curl} making the
right-hand side simply the Newtonian force. This implicit definition
for $\vecg_M$ could be inverted and used instead of $\vecg_N$ when
updating the particles' velocities during the time integration of
\Eq{eq:eom}. As noted in \Eq{eq:curl} they neglected the so-called
MONDian curl-field $\mathbf C=\nabla \times \mathbf h$. Others have
shown that this field decreases like $O(r^{-3})$ and vanishes for any
kind of symmetry \citep{Bekenstein84}, it yet remains unclear whether
it will leave any imprint on inhomogeneous strucure formation as found
in cosmological simulations.

In the second part of this paper we therefore presented a series of
cosmological simulations set out to quantify both structure formation
under MOND as well as the effects of the curl-field.  Surprisingly, we
found that our results are consistent with KG04 even if we take in
account that they use a different MOND model and initial conditions
generated with standard linear theory in a MONDian regimen.  We
revised some of their findings (mainly the discrepancy between the
abundance of galaxies at galaxies at high redshift between \LCDM\ and
OCBMond) and noted that the curl-field leaves marginal yet noticable
effects. The major result of this study though is that the curl-field
appears to drive structure formation! Our results obtained with the
new solver (and hence including the effects of the curl-field) can be
summarized as follows

\begin{itemize}
 \item the curl-field drives structure formation,
 \item the curl-field leads to more objects at $z=0$ where
 \item cross-identified objects are more massive in OCBMond2 than in OCBMond, and
 \item the OCBMond2 model shows a stronger two-point correlation function.
\end{itemize}

We acknowledge that there are still a lot of puzzling results to be
investigated in greater detail. However, we postpone this to future
studies. The aim of this paper was primarily to describe the novel
gravity solver that is freely available for download.\footnote{\amiga\
  can be downloaded from the following web page
  \texttt{http://www.aip.de/People/aknebe/AMIGA}. The new MOND solver
  can be switched on via the compilation flag \texttt{-DMOND2}.
The code is able to run cosmological simulations, ie with expansion and periodic boundary conditions and to work also in isolation, wich implies no expansion and fixed boundary conditions.  It can be also use without temporal evolution in order to get 3D potentials from density distributions given by particles or analitic formulas. 
Please contact the authors for
  more details.}


\section*{Acknowledgements}
CLL and AK acknowledge funding by the DFG under grant KN 755/2. AK
further acknowledges funding through the Emmy Noether programme of the
DFG (KN 755/1).  CLL is gratefull to Xufen Wu for her support during
the testing phase of the code.  The simulations and analysis were
carried out on the AIP's supercomputing cluster.

\bibliography{references} \bsp

\label{lastpage}

\end{document}